%% file: 00.Main.tex
\documentclass[sigconf]{acmart}
\usepackage{makecell}

\usepackage{float}            
\usepackage{placeins}         
\usepackage{array,booktabs}   
\usepackage{adjustbox}        
\usepackage{tabularx,longtable}
\usepackage[T1]{fontenc}
\usepackage{capt-of}          

\AtBeginDocument{%
  }

\setcopyright{acmlicensed}
\copyrightyear{2025}
\acmYear{2025}
\acmDOI{10.1145/XXXXXXX.XXXXXXX}
\acmConference[MSR 2026]{MSR '26: Proceedings of the 23rd International Conference on Mining Software Repositories}{April 2026}{Rio de Janeiro, Brazil}
\acmBooktitle{Proceedings of MSR '26: 23rd International Conference on Mining Software Repositories (MSR 2026)}
\acmISBN{978-1-4503-XXXX-X/2018/06}

\copyrightyear{2026}
\acmYear{2026}
\setcopyright{cc}
\setcctype{by}
\acmConference[MSR '26]{23rd International Conference on Mining Software Repositories}{April 13--14, 2026}{Rio de Janeiro, Brazil}
\acmBooktitle{23rd International Conference on Mining Software Repositories (MSR '26), April 13--14, 2026, Rio de Janeiro, Brazil}
\acmPrice{}
\acmDOI{10.1145/3793302.3793311}
\acmISBN{979-8-4007-2474-9/2026/04}

\begin{document}

\title{ A Large-Scale Dataset of MCP Implementations on GitHub}

\author{Benny Toeppe}
\affiliation{%
  \institution{Oakland University}
  \city{Michigan}
  \country{USA}
}
\email{btoeppe@oakland.edu}

\author{Amine Barrak}
\affiliation{%
  \institution{Oakland University}
  \city{Michigan}
  \country{USA}
}
\email{aminebarrak@oakland.edu}

\author{Emna Ksontini}
\affiliation{%
  \institution{University of North Carolina Wilmington}
  \city{North Carolina}
  \country{USA}
}
\email{ksontinie@uncw.edu}


\input{01.Abstract}




\maketitle

\input{02.Introduction}
\input{04.design-study}

\input{05.Dataset-Description}
\input{07.Threats}
\input{09.Conclusion}

\bibliographystyle{ACM-Reference-Format}
\bibliography{references}
\end{document}

%% file: 01.Abstract.tex
\begin{abstract}
The rapid emergence of the Model Context Protocol (MCP) has introduced a new standard for connecting large language models to external tools and services. Despite its rapid adoption in open-source development, systematic understanding of how MCP is implemented, structured, and maintained remains limited. This study presents the first large-scale, evidence-based dataset of real-world MCP implementation collected directly from GitHub. Using a hybrid pipeline that integrates the GitHub REST and GraphQL APIs with custom Python verification scripts, 3,238 candidate repositories were discovered, filtered, and validated through multi-stage evidence checks. Each verified project was classified by operational role (e.g., client, server, gateway) and exported in a reproducible JSONL schema. A manual review of a representative subset confirmed an overall precision of 83\% at a 95\% confidence level, and additionally revealed a set of repositories functioning primarily as educational samples, tutorials, or demonstration templates. A targeted exclusion rule was then applied to remove these non-operational repositories, resulting in a final dataset of 2,297 validated MCP projects. The analysis shows that Python and TypeScript dominate MCP development, with hybrid architectures emerging as the most common design pattern. By emphasizing transparent verification strategies, structured evidence tagging, and reproducible data organization, this work establishes a foundational benchmark for studying real-world MCP ecosystems and supports future research on integration, connectivity, and compatibility across the broader developer community.
\end{abstract}

%% file: 02.Introduction.tex
\section{Introduction}
\label{sec:introduction}

Modern language model systems are moving from pure text generation to agents that use tools and live data, enabling them to plan tasks, call external services, and work with information that changes over time rather than relying only on training data \cite{raieli2025building, barrak2025traceability}. Early attempts to connect LLMs with external tools relied on bespoke, ad hoc solutions such as proprietary function-calling mechanisms and custom API integrations \cite{schick2023toolformer, tang2023toolalpaca}. These proved that models could call external functions but created significant and unsustainable friction for developers, since every new tool or data source required custom integration code that tightly coupled the tool to a specific model or host application \cite{mastouri2025making}. A more scalable approach is to let agents communicate with tools and data sources through a shared protocol. The Model Context Protocol is the most visible effort in this direction. It defines how an application with an embedded model discovers capabilities and invokes them through a simple client–server design, allowing any compliant client to discover and call any compliant server through a consistent interface \cite{ferrag2025llm}.

However, we still lack a clear, code grounded picture of how this protocol is implemented at scale. Public documentation and vendor posts explain the protocol and its goals, but they do not show which concrete design choices developers make in repositories, how those implementations evolve over time, or what evidence links a listed project to a working client or server. Early empirical work confirms why this gap matters. Hasan et al. \cite{hasan2025model} analyze 1,899 servers and report protocol specific vulnerabilities and maintainability concerns using static analysis and an MCP focused scanner, which underscores the need for datasets that let researchers connect ecosystem level claims to the actual code that developers publish. To date, the main attempts to map the ecosystem at scale rely on registry or marketplace listings rather than on repository centric verification. Lin et al. \cite{lin2025large} introduce MCPCorpus by starting from MCP.so and then enriching entries with GitHub level signals such as stars, forks, contributors, and last commit time. This market centered efforts are valuable for ecosystem measurement, yet they do not provide a repository first and verifiable ground truth of MCP implementations that is built from code evidence in the repositories themselves.

In this paper, we construct a large scale repository centered dataset of MCP implementations collected directly from GitHub. Our pipeline discovers candidate repositories with the GitHub REST and GraphQL APIs, then applies multi stage, evidence based verification to confirm that each project is a working server, client, or gateway. The verification uses tangible code level signals, including dependency fingerprints from package manifests and protocol specific file layouts, and it records these signals as evidence for each entry. A manual review over a representative sample confirms the precision of the pipeline, which supports reproducibility and transparent curation. The final dataset contains 2,297 validated projects out of 3,238 discovered candidates, organized in a reproducible JSONL schema. Beyond point in time metadata, we also collect change history to support evolution studies, including commit records and pull requests that modify MCP related files and configuration paths.

The complete dataset, including implementation evidence, classification outputs, and code, is publicly accessible at \cite{mcp_dataset_2025}.

%% file: 04.design-study.tex
\section{Dataset Construction}
\begin{figure}[h]
    \centering
    \includegraphics[width=0.9\linewidth]{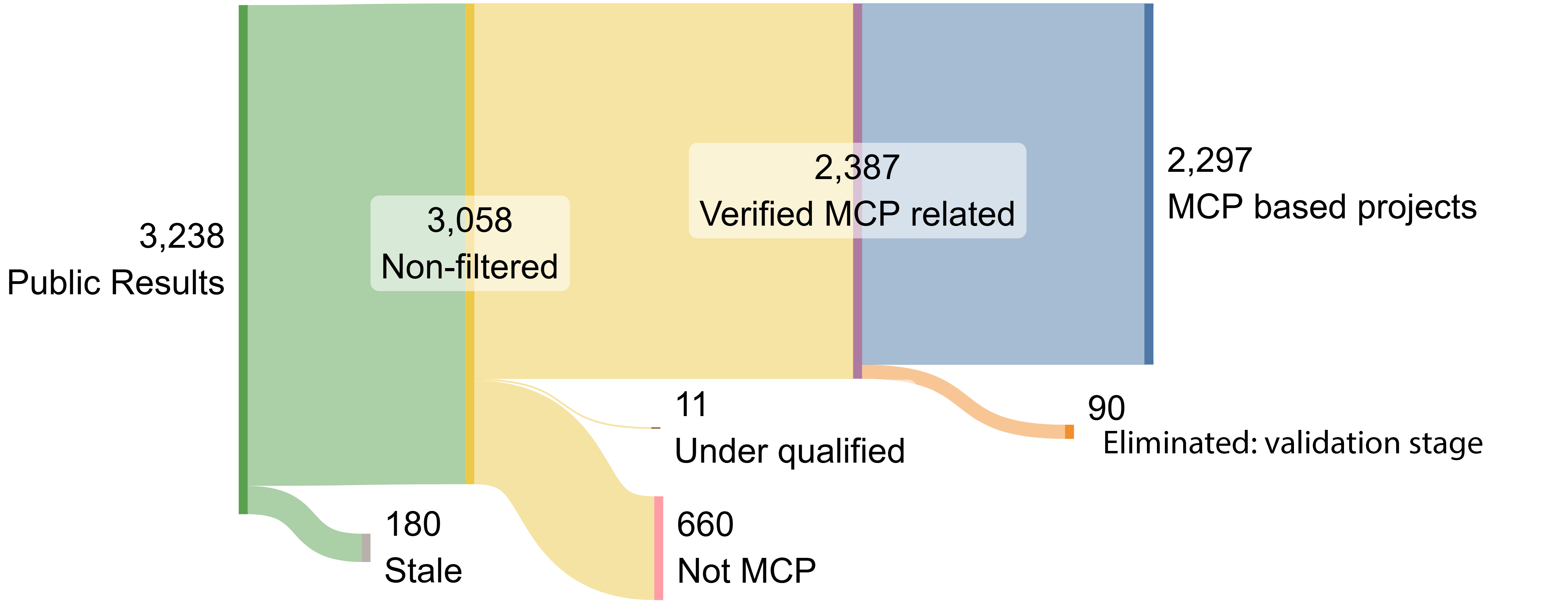}
    \vspace{-0.5em}
    \caption{Flow of dataset construction from initial GitHub results to the final verified MCP implementations.}
    \vspace{-2em}
    \label{fig:placeholder}
\end{figure}

Figure \ref{fig:placeholder} summarizes our dataset construction pipeline. We began with 3,238 public repositories referencing the Model Context Protocol (MCP). After removing stale and documentation-only repositories, 3,058 remained. Our verification pipeline confirmed MCP-related code artifacts in 2,387 repositories. A final refinement step based on manual validation removed 90 non-operational projects, resulting in 2,297 verified MCP implementations.

\subsection{Initial Repository Discovery}
We collected candidate repositories directly from GitHub using the REST Search API combined with GraphQL metadata retrieval. To ensure relevance to the emergence of MCP-driven agentic workflows, we limited the dataset to repositories created or updated between \textbf{January 2024} and \textbf{October 2025}.

We designed five search queries to capture both broad references to MCP and more specific implementation contexts. Two queries targeted explicit references to the protocol, while three focused on contexts where MCP usage is strongly associated with implementation, such as Claude Desktop integrations and server or client role naming. Table \ref{tab:search-queries} summarizes the total results returned from GitHub and the number retained after filtering for duplicates, false positives, and archived projects.



\begin{table}[htbp]
  \centering
  \small
  \renewcommand{\arraystretch}{1.15}
  \caption{GitHub search queries and resulting repository}
  \vspace{-0.5em}
  \label{tab:search-queries}

  \resizebox{\linewidth}{!}{%
  \begin{tabular}{|l|r|r|}
    \hline
    \textbf{Query} & \makecell{\textbf{Total}\\ \textbf{repositories}\\ \textbf{(response)}} & \makecell{\textbf{Total}\\ \textbf{repositories}\\ \textbf{(Kept)}} \\
    \hline
    \texttt{"MCP" in:name,readme,description} & 3{,}148 & 2{,}379\\
    \texttt{"mcp server" in:name,readme,description} & 2{,}410 & 1{,}999\\
    \texttt{"Model Context Protocol" in:name,readme,description} & 1{,}815 & 1{,}521\\
    \texttt{"Claude Desktop" MCP in:name,readme,description} & 1{,}093 & 979\\
    \texttt{"mcp client" in:name,readme,description} & 800 & 743\\
    \hline
  \end{tabular}%
  }

  \vspace{-1em}
\end{table}

After merging overlapping results across queries and removing duplicates, we obtained 3,238 unique repositories.

\subsection{Filtering for Active Maintenance}

To ensure that the dataset reflects active MCP development rather than abandoned prototypes or documentation, we applied the following filters:

\begin{itemize}
    \item Repositories with no commits within the last nine months were removed.
    \item Repositories consisting solely of documentation, tutorials, or announcement stubs were removed.
    \item Forks without modification beyond the upstream version were excluded.
\end{itemize}

This resulted in a refined candidate set of 3,058 repositories.

\subsection{Multi-Stage Verification Pipeline}
We applied a verification pipeline designed to detect operational MCP implementations. Each repository was examined across three evidence layers:

\subsubsection{Manifest and Dependency Evidence}
We searched ecosystem-specific manifest and lock files to identify MCP-related dependencies, including:

\begin{itemize}
    \item JavaScript: \texttt{package.json}, \texttt{package-lock.json}, \texttt{yarn.lock}
    \item Python: \texttt{pyproject.toml}, \texttt{Pipfile.lock}, \texttt{poetry.lock}
    \item Rust, Go, PHP: \texttt{Cargo.toml}, \texttt{go.mod}, \texttt{composer.json}
\end{itemize}

Common dependency indicators included:
\texttt{mcp/sdk}, \texttt{fastmcp}, \texttt{mcp}, and \texttt{mcp-client}.

\subsubsection{Directory Structure and Entry Points}

Because manifests alone do not ensure operational implementation, we analyzed:

\begin{itemize}
    \item Directory patterns such as \texttt{/src/mcp/}, /cmd/mcp/, and /bin/mcp-server.js
    \item Executable entry points configured via script definitions (e.g., \texttt{"mcp:run"}, \texttt{mcp-server}, \texttt{[project.scripts]})
\end{itemize}

Repositories exhibiting MCP entry points were marked as executable MCP systems.

\subsubsection{Integration and Environment Evidence}

We inspected deployment environments via:

\begin{itemize}
    \item CI/CD workflows (e.g., \texttt{.github/workflows})
    \item Containerization files (e.g., \texttt{Dockerfile}, \texttt{docker-compose})
    \item Claude Desktop configuration files (e.g., \texttt{config.json})
\end{itemize}

This step identified 660 repositories with no operational MCP artifacts, which were removed, along with 11 incomplete implementation scaffolds. This resulted in 2,387 verified MCP repositories.



\subsection{Manual Validation and Rule Refinement}
To evaluate the automated verification stage, two authors independently reviewed the 170-repository subset, which was selected at the 95\% confidence level with an estimated margin of error of ±7–8\%. Each author examined whether the project implemented MCP by inspecting configuration files, dependency declarations, and entry-point code. Disagreements were resolved through discussion to reach a final consensus, reducing individual bias. This validation revealed several recurring sources of false positives:

\begin{itemize}
\item Repositories using \texttt{"MCP"} to refer to unrelated systems (e.g., Minecraft Coder Pack).
\item Reference-only mentions of MCP without executable server or client code.
\item Template repositories with placeholder code and no functional implementation.
\end{itemize}

The manual assessment between the two authors found that 83.0\% of the reviewed repositories represented genuine MCP implementations, while 12.0\% were false positives and 5.0\% were false negatives. Most errors stemmed from language-specific file conventions or incomplete dependency manifests. Extrapolating these proportions to the initial set of 2,387 automatically verified repositories yields a Wilson confidence interval of [79.0\%, 87.0\%] at the 95\% confidence level.
We refined the filtering rules based on these findings, removing 90 additional false matches and producing a final dataset of 2,297 verified MCP implementations.

%% file: 05.Dataset-Description.tex
\section{Dataset Contents and Characteristics}
This section outlines the 2,297 verified MCP repositories, the operational roles they implement, how roles interact, and how role assignments and metadata were derived.
\subsection{Role-Based MCP Execution Model}
\begin{figure}[h!]
    \centering
    \includegraphics[width=0.9\linewidth]{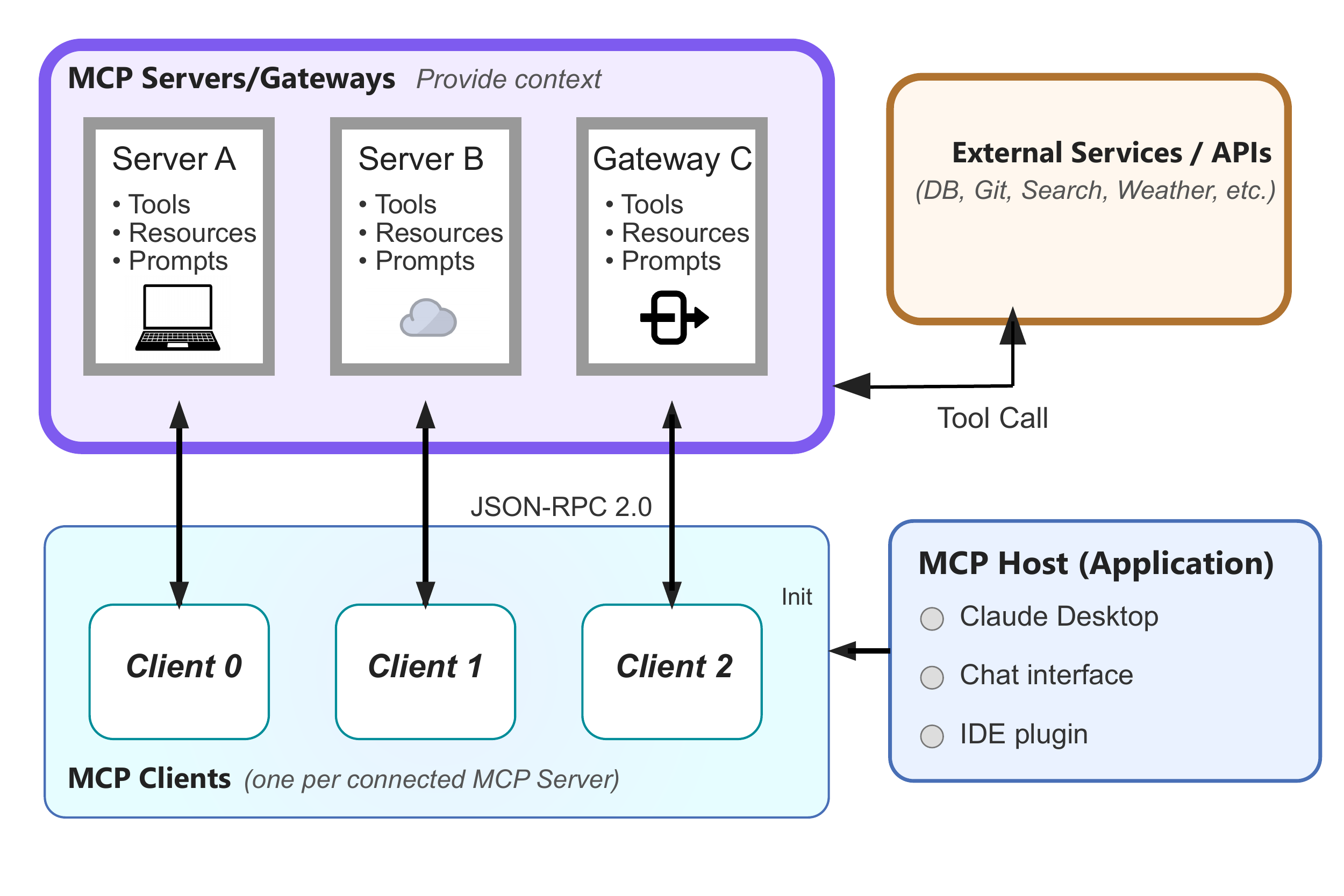}
    \vspace{-1.5em}
    \caption{MCP architecture showing the host, clients, and servers/gateways exposing tools and resources.}
    \label{fig:mcp-roles}
    \vspace{-1.5em}
\end{figure}

The Model Context Protocol defines how agents invoke external tools through a structured request-response workflow. In practice, implementations of MCP adopt one of three primary operational roles that correspond to different points in the execution chain. Figure~\ref{fig:mcp-roles} illustrates the interaction pattern among these roles.


\begin{itemize}
    \item \textbf{Client}: initiates connections to MCP servers and consumes their tools, resources, or prompts. Typical examples include user-facing interfaces such as Claude Desktop integrations.
    
    \item \textbf{Server}: Exposes MCP-compatible tools, resources, or prompts that Clients can invoke. Servers implement the logic for computation, retrieval, transformation, or control when handling client requests.
    
    \item \textbf{Gateway}: Mediates MCP traffic between components or external ecosystems. Gateways translate, route, or relay requests, often bridging MCP with APIs.
\end{itemize}

\subsection{Role Classification Rules}

Each verified repository was assigned a primary operational role based on source-level evidence. The role classification used a rule-based scoring system combining manifest signals, entry-point patterns, and semantic cues from metadata and documentation, resulting in one of three roles: \textbf{Client}, \textbf{Server}, or \textbf{Gateway}.

\subsubsection*{\underline{Rule-Based Scoring System}}

Three independent score variables were computed for each repository:
$S_c$ (client score), $S_s$ (server score), and $S_g$ (gateway score). Signals were drawn from repository metadata, directory structure, source paths, dependency manifests, and commit-level evidence.

\begin{itemize}
    \item \textbf{Client signals}: Use of MCP client libraries and explicit tool invocation (e.g., \texttt{connect()}, \texttt{invoke()}, \texttt{use\_tool()}, paths such as \texttt{@mcp/sdk/client}), including editor plugin contexts (e.g., \texttt{vscode}, \texttt{cursor}, \texttt{claude\_desktop}).

    \item \textbf{Server signals}: Registration and exposure of MCP tools or capabilities, indicated by directories like \texttt{/tools/} or /resources/, manifest references to \texttt{@mcp/sdk} or \texttt{fastmcp}, and server entrypoints such as \texttt{server.py}, \texttt{server.ts}, or mcp-server binaries.

    \item \textbf{Gateway signals}: Logic forwarding MCP requests across systems, indicated by terms such as \texttt{gateway}, \texttt{router}, \texttt{relay}, or \texttt{proxy}, and bridges to external APIs or services. Observed patterns fall into six subtypes: transport relay, language bridge, authentication policy, orchestration, sandbox isolation, and discovery/registry.

\end{itemize}

Each match contributed to a numeric score based on indicator strength (strong = 2 points, weak = 1 point), where strong signals derived from source and manifest files and weak signals from README/metadata text.

\subsubsection*{\underline{Decision Logic}}

The final role was determined based on the comparative magnitude of the three scores:

\begin{enumerate}
    \item If $S_c \geq 3$ and $S_s \leq 1$, classify as \textbf{Client}.
    \item If $S_s \geq 3$ and $S_c \leq 1$, classify as \textbf{Server}.
    \item If $S_g \geq 4$ and both $S_c$ and $S_s < 2$, classify as \textbf{Gateway}.
\end{enumerate}

A \textbf{client override rule} was applied to prevent misclassification of editor extensions and IDE integrations that contain incidental routing logic. If gateway-scored repositories contained editor signals (e.g., \texttt{vscode}, \texttt{cursor}, \texttt{claude\_desktop}), the final classification was reassigned to \textbf{Client}.

\begin{figure*}[h!]
    \centering
    \includegraphics[width=\linewidth]{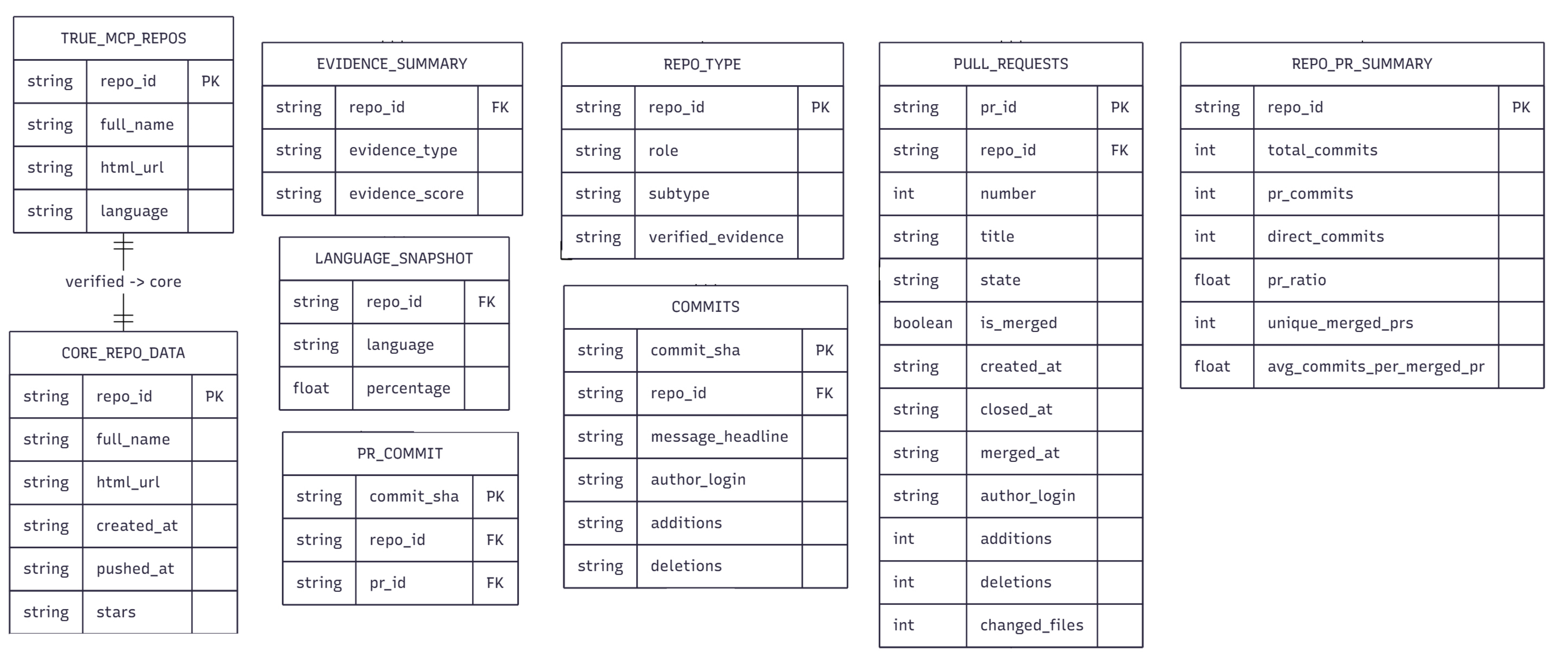}
    \caption{Dataset Schema and Entity Relationships for the MCP Repository Dataset}
    \label{fig:dataset-schema}
    \vspace{-1em}
\end{figure*}

\subsubsection*{\underline{Manual Disambiguation}}

Repositories where the scoring system did not yield a clear dominant role (e.g., when $|S_c - S_s| \leq 1$ or when all three scores were below threshold) were marked as \textit{low confidence}. These repositories were reviewed manually to determine whether they primarily issued tool requests, exposed tools, or mediated between systems. If a dominant operational role could not be identified after inspection, the repository was labeled as \textbf{Unclassified}.

Applying these rules resulted in 1,962 Servers, 1,462 Clients, and 80 Gateways. Some repositories exhibited both client and server behavior and were assigned to both roles, while 36 could not be assigned a dominant role and were marked as \textbf{Unclassified}.

Among the 80 Gateways, orchestration gateways were the most common (n=47), followed by language bridges (n=22). Transport relays (n=6) and authentication policy gateways (n=5) appeared less frequently, with sandboxing (n=2) and discovery/registry gateways (n=1) being rare.

The 36 Unclassified repositories were mostly lightweight or incomplete projects, including agent adapters (n=13), chat UI integrations (n=5), documentation or specification scaffolds (n=4), SDK utilities and CLI tooling (n=6), small demo or benchmark artifacts (n=4), and unstructured repositories (n=4).

\subsection{Language Characteristics by Role}
We examined the primary implementation language of each repository after role classification. Table~\ref{tab:lang_rank} presents the top three languages for each role, ranked by number of repositories.

\begin{table}[htbp]
\centering
\caption{Top 3 Languages per MCP Role}
\vspace{-0.5em}
\label{tab:lang_rank}
\begin{tabular}{|c|c|c|c|}
\hline
\textbf{Rank} & \textbf{Server} & \textbf{Client} & \textbf{Gateway} \\
\hline
1 & Python (785) & Python (530) & TypeScript (25) \\
2 & TypeScript (597) & TypeScript (492) & Python (22) \\
3 & JavaScript (177) & JavaScript (137) & Go (17) \\
\hline
\end{tabular}
\vspace{-0.25em}
\end{table}

Python and TypeScript dominate both server and client implementations. JavaScript appears in smaller client- and server-side projects and examples. Gateways show a distinct profile, with Go appearing as a third major language.

\subsection{Dataset Schema and Repository Metadata}

Figure~\ref{fig:dataset-schema} shows the schema used to store the verified MCP repositories and their metadata. The dataset centers on a core repository index, with auxiliary tables for language composition, role classification, evidence signals, and pull request activity.

\subsubsection*{\underline{Repository Core and Verification Evidence}}

The dataset centers on the \texttt{TRUE\_MCP\_REPOS} table, which lists the 2,297 verified repositories. Each repository is linked to:
\begin{itemize}
    \item \texttt{CORE\_REPO\_DATA}, containing metadata such as creation date, last push timestamp, and star count.
    \item \texttt{EVIDENCE\_SUMMARY}, which records the types of signals that contributed to verification (manifest, entry points, directory patterns, etc.) along with evidence strength scores.
    \item \texttt{REPO\_TYPE}, which defines the repository’s assigned role (Client, Server, Gateway), subtype where applicable, and supporting evidence.
\end{itemize}


\subsubsection*{\underline{Commit-Level Activity}}
Development history is stored in the \texttt{COMMITS} table, which records commit messages, author identity, and code change size. Across all verified repositories, we observe:

\begin{itemize}
\item Median commits per repository: \textbf{70}
\item Median commit message length: \textbf{38 characters}
\item Median additions per commit: \textbf{19 lines}
\item Median deletions per commit: \textbf{4 lines}
\item Median changed files per commit: \textbf{2 files}
\item Median number of unique contributors per repository: \textbf{4}
\item Median repository stars: \textbf{156}
\item Median forks per repository: \textbf{27}
\item Median open issues per repository: \textbf{3}
\item Median number of published releases: \textbf{1}
\end{itemize}

These patterns suggest iterative, small-step development styles rather than large batch updates.

\subsubsection*{\underline{Pull Request Structure and Collaboration}}
Pull request metadata is stored in the \texttt{PULL\_REQUESTS} table, with aggregate summaries in \texttt{REPO\_PR\_SUMMARY}. We track counts of merged PRs, review patterns, and author involvement. At the dataset level, we observe:

\begin{itemize}
    \item Most MCP servers and clients are maintained by \textbf{small teams} (2--5 active contributors).
    \item Gateway repositories show \textbf{higher PR activity}, with an average of 49.6 pull requests per repository, compared to 33.7 for clients and 29.9 for servers.

\end{itemize}

%% file: 09.Conclusion.tex
\section{Conclusion}
\label{sec:conclusion}
We constructed a dataset of 2,297 verified MCP repositories using a multi-stage discovery and verification process based on source-level evidence. The dataset records role assignments, evidence indicators, language composition, and development activity in a structured format. Python and TypeScript are the most common languages in both client and server implementations, while gateways additionally use Go. Several repositories include both client and server functionality within the same codebase. The dataset is intended to support reproducible analysis of MCP implementation practices, with future work focusing on updating the dataset over time and refining role and subtype classifications.



